\documentclass[preprint2]{emulateapj}

\begin{document}

\title{WISE Detection of the Galactic Low-Mass X-Ray Binaries}

\author{Xuebing Wang\altaffilmark{1} and Zhongxiang Wang}

\affil{Key Laboratory for Research in Galaxies and Cosmology, 
Shanghai Astronomical Observatory,\\
Chinese Academy of Sciences, 80 Nandan Road, Shanghai 200030, China}

\altaffiltext{1}{\footnotesize Graduate University of Chinese Academy of 
Sciences, No. 19A, Yuquan Road, Beijing 100049, China}

\begin{abstract}
We report on the results from our search for the Wide-field Infrared Survey 
Explorer detection of the Galactic low-mass X-ray binaries. Among 187 binaries
catalogued in Liu et al. (2007), we find 13 counterparts and two candidate
counterparts. For the 13 counterparts, two (4U~0614+091 and GX~339$-$4) 
have already been confirmed by previous studies  
to have a jet and one (GRS~1915+105) to have a candidate circumbinary 
disk, from which the detected infrared emission arose. 
Having collected the broad-band optical and near-infrared data 
in literature and constructed flux density spectra for the other 10 binaries, 
we identify that three (A0620$-$00, XTE J1118+480, and GX 1+4) are candidate 
circumbinary disk systems, four (Cen X-4, 4U 1700+24, 3A 1954+319, and Cyg X-2)
had thermal emission from their companion stars, and 
three (Sco X-1, Her X-1, and Swift J1753.5$-$0127) are peculiar systems 
with the origin of their infrared emission rather uncertain. 
We discuss the results and WISE counterparts' brightness distribution 
among the known LMXBs, and suggest that more than half of the LMXBs would 
have a jet, a circumbinary disk, or the both.

\end{abstract}


\section{INTRODUCTION}

X-ray binaries (XRBs), containing either an accreting neutron star or
black hole,
constitute a large fraction of bright X-ray sources in the Galaxy.
When they have $\lesssim 1~M_{\sun}$ low mass companions, a companion overfills
its Roche lobe, and mass transfer from the companion
to the central compact star occurs via an accretion disk. Such binaries are
further categorized as low-mass X-ray binaries (LMXBs). Besides their prominent
X-ray emission, LMXBs are generally observable at optical
wavelengths, as the accretion disks and/or companion stars radiate
sufficiently bright thermal emission. In addition, the LMXBs
are known to be able to launch a jet, producing synchrotron emission
detectable from optical/infrared to radio wavelengths
(e.g., \citealt{fen06, rus+06, rus+07, gal10}). Studies of jets  help
understand flux variabilities seen in LMXBs, constrain the fractions of
accretion energy channeled to different emission components, and thus allow
to draw a full picture for the detailed physical processes occurring in
LMXBs.

The environments in which XRBs are located may be dusty. The supernova
explosions that produced the compact stars are thought to have a fallback
process \citep{che89}, due to the impact of the reverse shock wave with
the outer stellar envelope. As a result, part of the ejected material
during a supernova explosion may fallback to the newly born compact star.
During the evolution of a LMXB, a substantial amount of mass has been lost
from the companion (e.g., \citealt{prp02}). Observational evidence as well
as theoretical studies show that disk winds or outflows are probably
ubiquitous and may be massive (e.g., \citealt{nei13, ybw12} and references
therein). If a small fraction of the material from any of the processes
is captured to be around
a binary, it is conceivable that a circumbinary disk might have formed from
the material, acting to intercept
part of emission (X-rays from the central compact star and optical light
from the companion and accretion disk) from the binary system and
re-radiate the energy at infrared (IR) wavelengths. 
Indeed, \citet{mm06} observed
four LMXBs with \textit{Spitzer} Space Telescope and found that two of them
might harbor a circumbinary dust disk. Moreover, the \textit{Spitzer} 
detection of dust emission features in a so-called microquasar
(see, e.g., \citealt{gal10}) clearly indicates the existence of dust material
around the LMXB \citep{rah+10}.

The Wide-field Infrared Survey Explorer (WISE),
launched in 2009 December, mapped the whole sky at its bands of
3.4, 4.6, 12, and 22 $\mu$m (called W1, W2, W3, and W4, respectively) in 2010
\citep{wri+10}. The FWHMs of the averaged point spread function for WISE
imaging at the four bands
were 6\farcs1, 6\farcs4, 6\farcs5, and 12\farcs0, and the sensitivities
(5$\sigma$) generally reached were 0.08, 0.11, 1, and 6 mJy.
In the WISE all-sky images and source catalogue released in 2012 March,
measurements of over 563 million objects are provided.
Therefore WISE
imaging has provided sensitive measurements of many different types of
celestial objects at the IR wavelengths (see \citealt{wri+10} for details).

Using the WISE data, we have carried out searches for the counterparts
to the LMXBs catalogued in \citet{lvv07}, for the purpose of identifying 
sources among them with either jets or circumbinary debris disks. 
In this paper we report the results from our searches.
\begin{figure}
\centering
\includegraphics[scale=0.58]{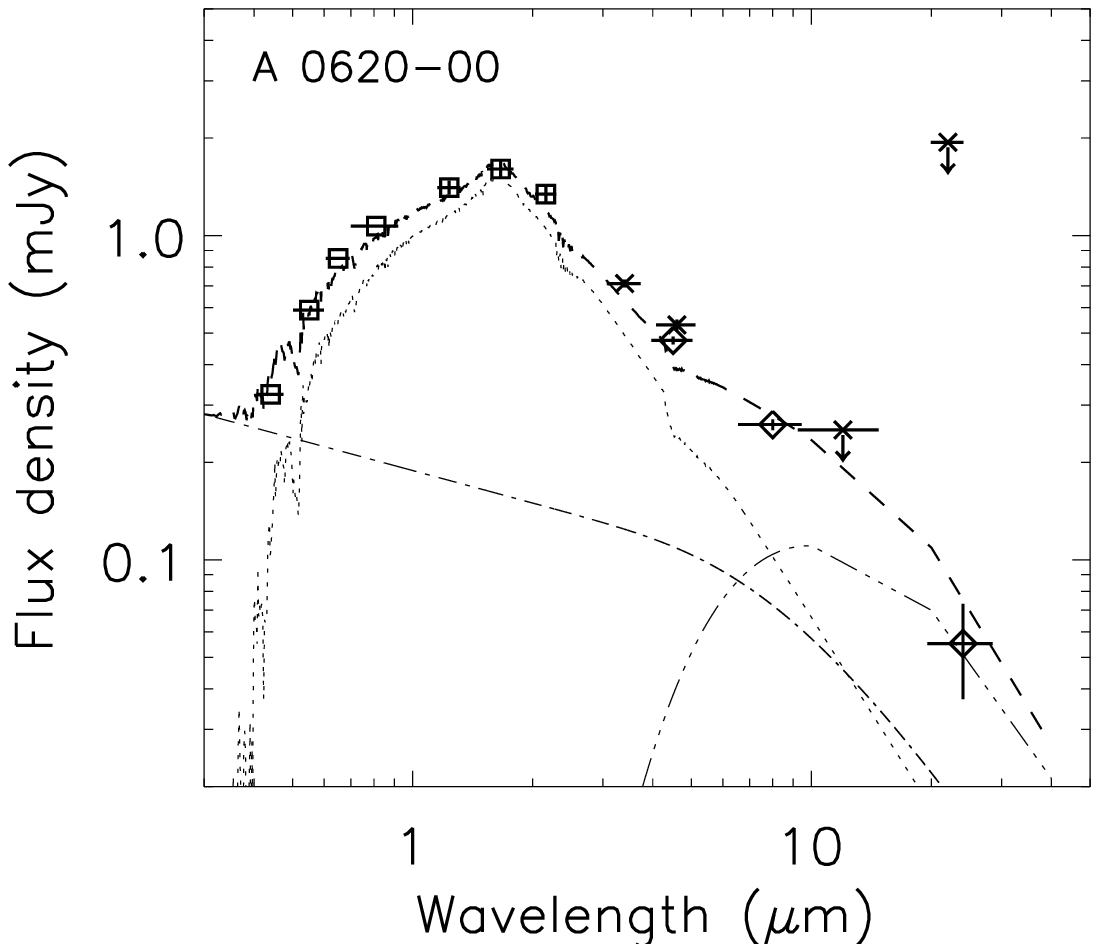}
\includegraphics[scale=0.58]{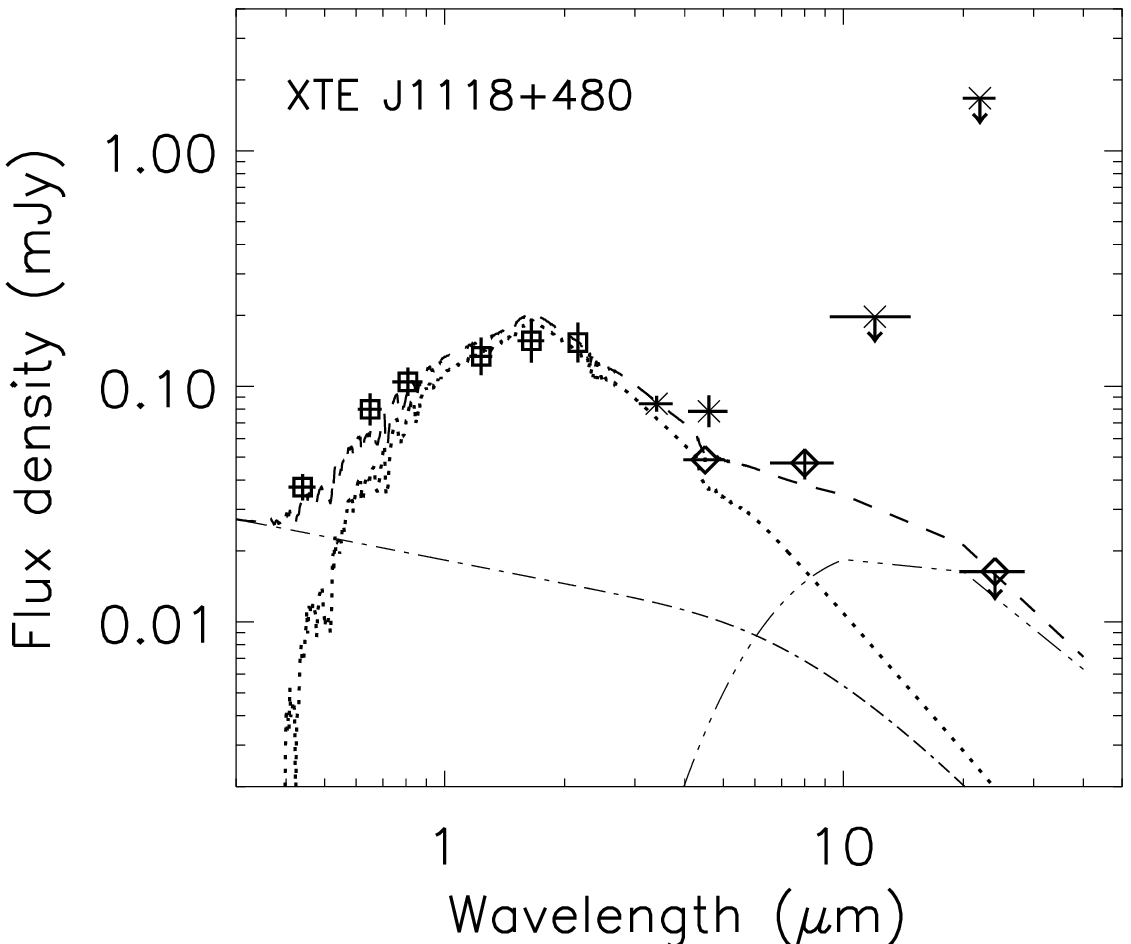}
\includegraphics[scale=0.58]{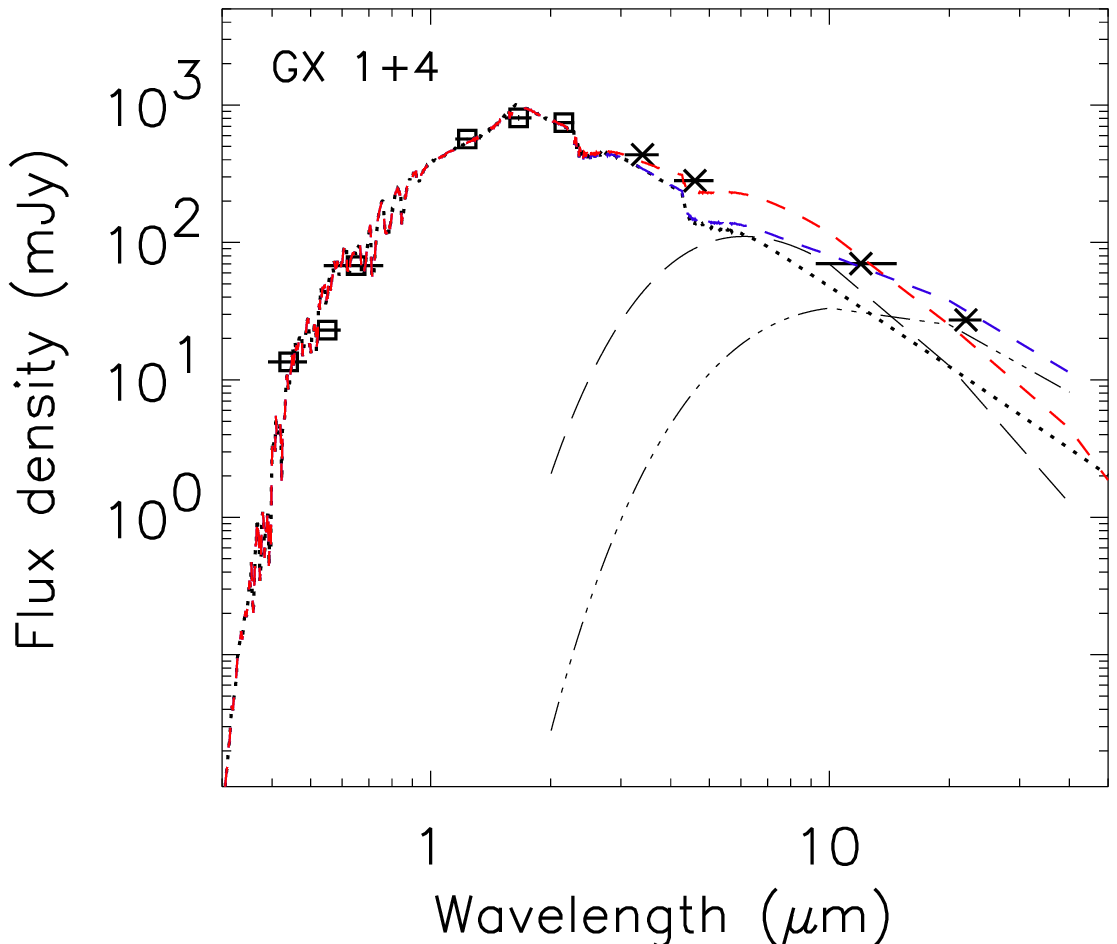}
\caption{Flux density spectra of A0620$-$00, XTE J1118+480, and GX~1+4.
The squares and diamonds are optical/near-IR and \textit{Spitzer} data points,
respectively, and crosses are the WISE data points. The dotted, dash-dotted,
and dash--triple-dotted curves represent emission from the companions, 
accretion disks, and circumbinary disks, respectively. The dashed curves
(blue one for GX 1+4)
represent the total emission from all the components.
For GX 1+4, the long dashed curve represents emission from a dust shell, and
a red dashed curve is the total emission from the companion and the dust shell.
\label{fig:bd}}
\end{figure}

\section{DATA ANALYSIS}
\label{sec:daa}

The LMXB catalogue contains 187 sources in the Galaxy, Large and Small
Magellanic Cloud \citep{lvv07}. The general properties of these sources,
for example their coordinates, magnitudes in the $UBV$ bands,
estimates of the interstellar reddening $E(B-V)$, X-ray types,
are provided when such information is available in literatures .
The X-ray types such as X-ray burst source, X-ray pulsar, or microquasar,
allow to know whether the primary compact object is a black hole
or a neutron star.

We input the coordinates of the LMXBs given in the catalogue into
WISE Source Catalog query, provided at the NASA/IPAC Infrared Science Archive.
Given the FWHMs of WISE imaging, we used a 2\farcs0 radius 
(see \citealt{deb+11} for details) around the coordinates for counterpart 
searching.  All sources located in globular clusters
were excluded because of the high density of stars in a cluster.
Candidate counterparts to 17 LMXBs were found.
For them, we further checked the uncertainties for their reported source
positions, and found that most of them have positional uncertainties of
$\lesssim 1\arcsec$, sufficiently small comparing to the 2\arcsec\
search radius.
However among them, the position of 4U~1735$-$28 has an
uncertainty of 7\arcmin\ \citep{for+78}, too large for finding
the counterpart, and it was excluded. In addition for
SLX 1737$-$282, the position in the LMXB catalogue was from
\citet{int+02}, which has an uncertainty of 8\farcs3. A \textit{Chandra}
position was reported by \citet{tom+08}, which has an uncertainty of 0\farcs6
and an offset of 3\farcs3 from the WISE source. The WISE source was
detected by 2MASS \citep{2mass} at $JHK_s$ bands, which was already 
excluded as the counterpart to SLX~1737$-$282 by \citet{tom+08}.
Further checking the previous optical/IR studies of each of the sources, 
we concluded that counterparts to 13 LMXBs and candidate 
counterparts to 2 LMXBs were found.
The offsets of the WISE source coordinates from the input ones of the
15 LMXBs are summarized in Table~\ref{tab:sum}.
The WISE magnitude measurements or upper limits of them, as well as
2MASS $JHK_s$ measurements if they were detected, are also given in
Table~\ref{tab:sum}.

To determine any excess emission in the IR bands, thermal emission
from a companion star has to be often considered.
On the basis of the reported spectral types for the companion stars,
we used stellar atmosphere models of the
\citet{kur93} to estimate the thermal emission.
If any flux measurements at optical and near-IR bands
for a source are available, the information was collected and then 
used to construct a broad-band spectrum, which is required for 
the determination of the presence of any excess emission.
The reddening values to the LMXBs were also collected in order to deredden
the observed fluxes. 
The reddening laws of \citet*{sfd98} for the optical and near-IR data,
and \citet{ind+05} (wavelengths $\leq$8 $\mu$m) and \citet{wd01}
(wavelengths $>$8 $\mu$m) for \textit{Spitzer} (when included) and
WISE data were used.
The information of the binary types, orbital periods,
distance estimates from observations, optical photometric bands,
and reddening values for the 15 LMXBs, which was used for the construction
and analysis of broad-band spectra, is summarized in Table~\ref{tab:prop}.

\section{Results}
\label{sec:res}

\subsection{Candidate circumbinary debris disk systems}

On the basis of their broad-band spectra and previous multi-wavelength
studies, we identified four LMXBs with a candidate circumbinary debris disk.
They are A0620$-$00, XTE J1118+480, GX 1+4, and GRS 1915+105.
For the last source, while it is a well-known X-ray binary with a 
jet \citep{fb04}, multiple \textit{Spitzer} mid-IR spectroscopic 
observations have detected PAH features in its emission, clearly indicating the
existence of (possibly) a circumbinary dust disk \citep{rah+10}.
Comparing the \textit{Spitzer} flux measurements given by 
\citet{rah+10} with that from WISE, the flux values are comparable given
the source is highly variable. We refer to \citet{rah+10} for detailed models 
built to explain the mid-IR emission.
\begin{figure*}
\begin{center}
\begin{tabular}{ll}
\includegraphics[scale=0.58]{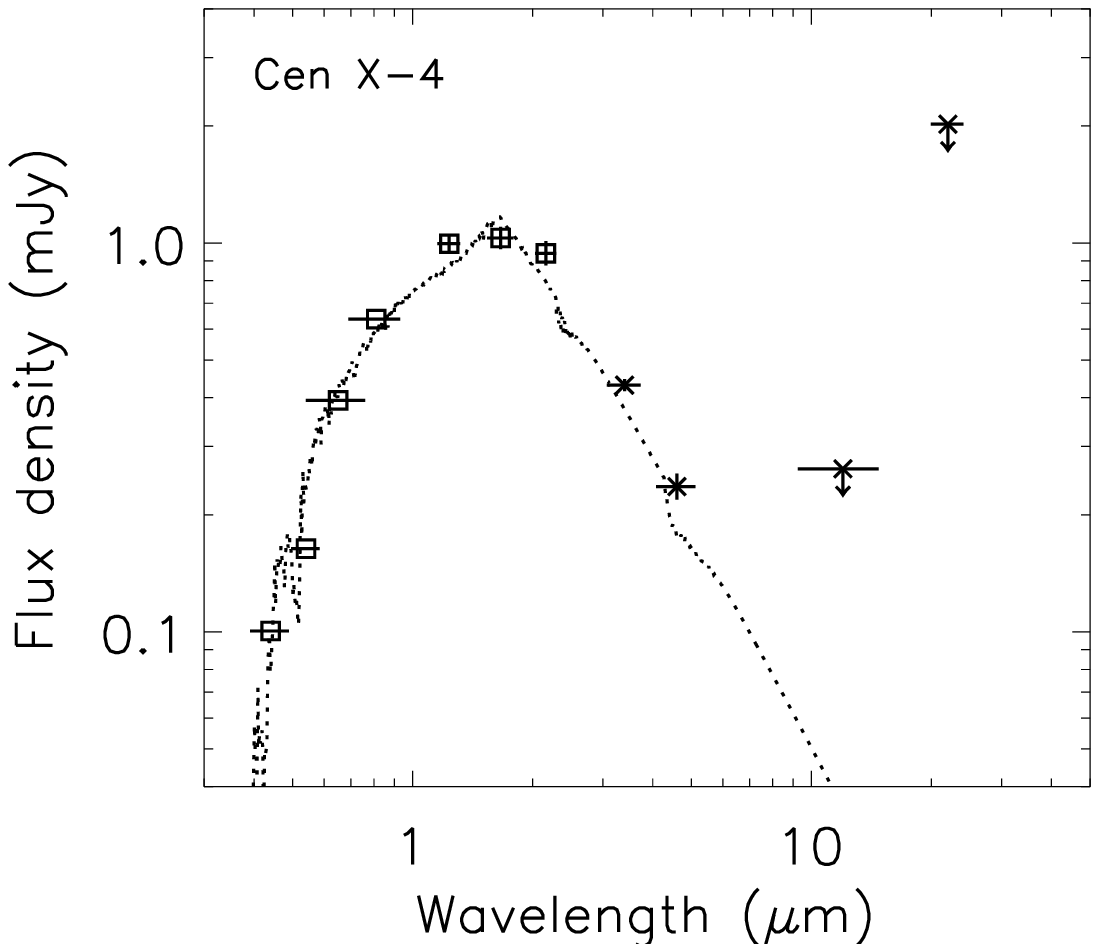}
\includegraphics[scale=0.58]{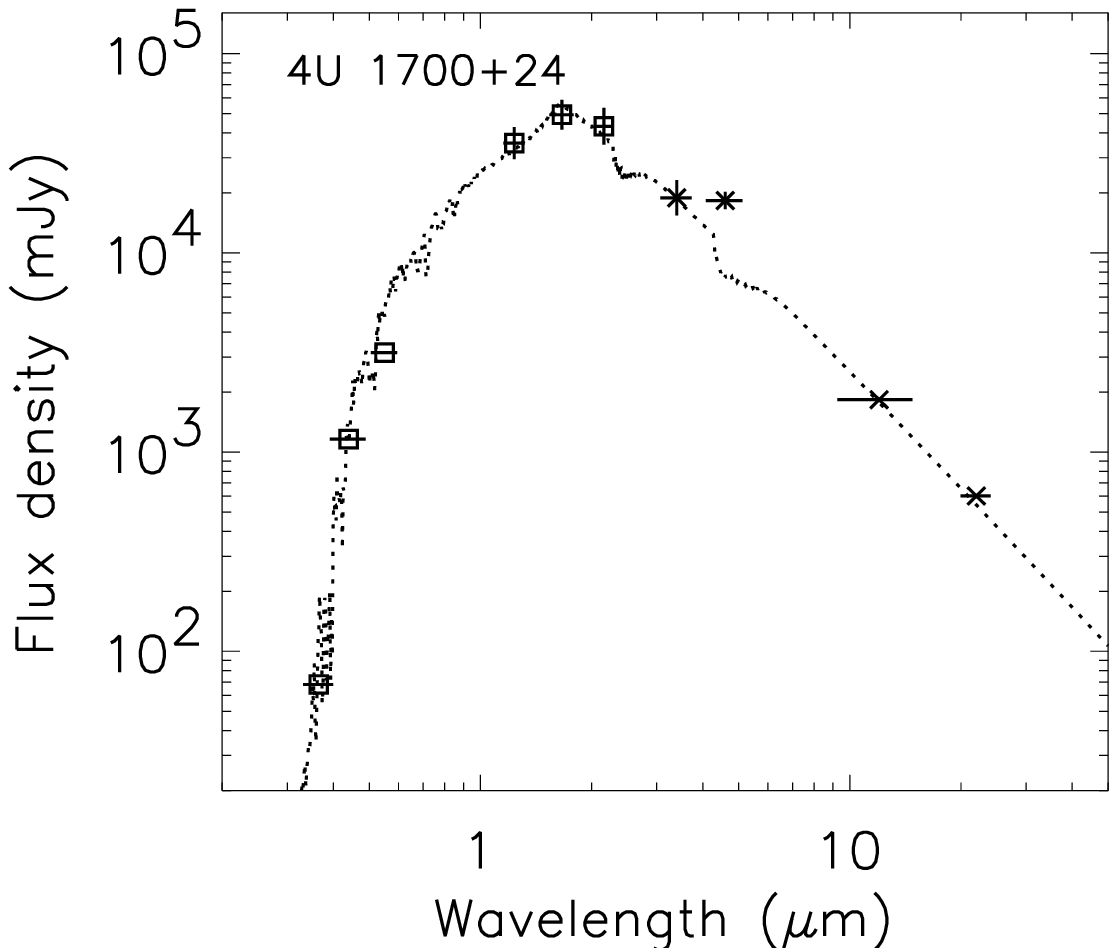}\\
\includegraphics[scale=0.58]{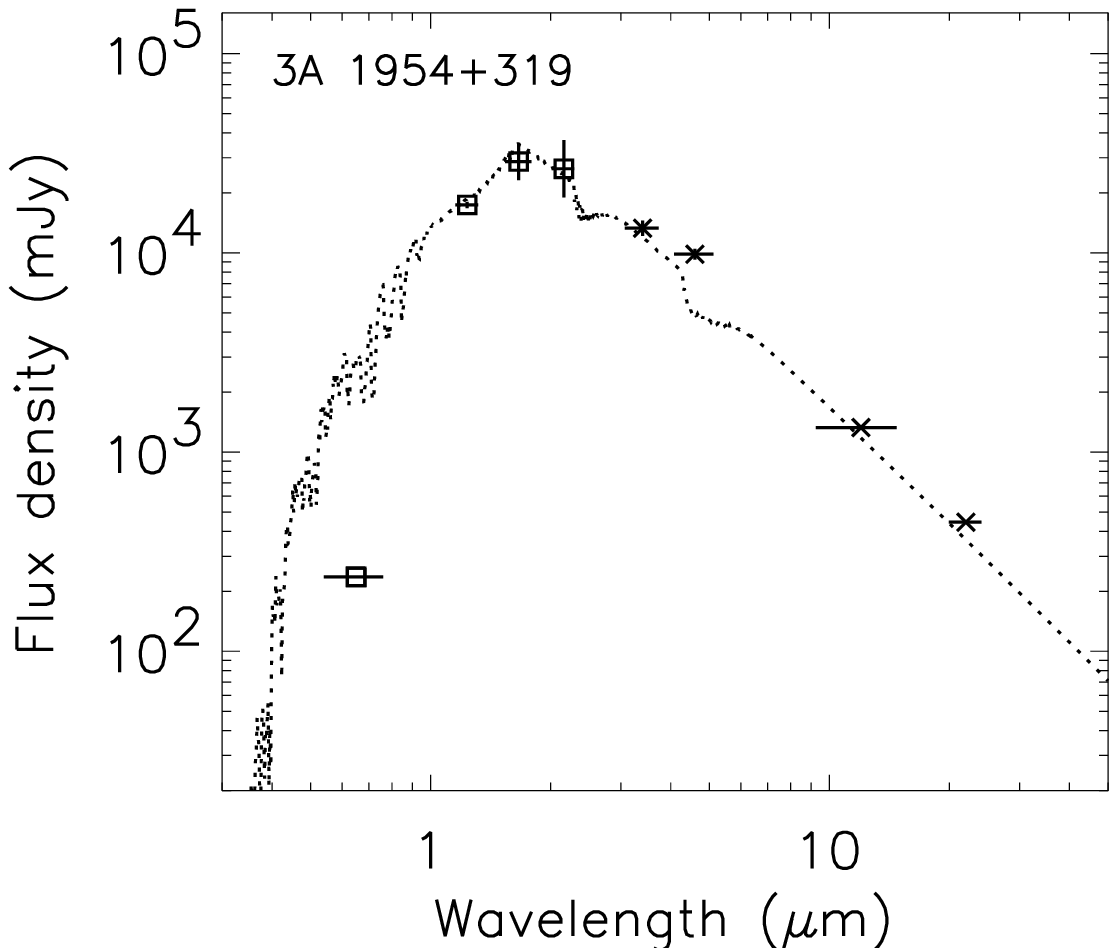}
\includegraphics[scale=0.58]{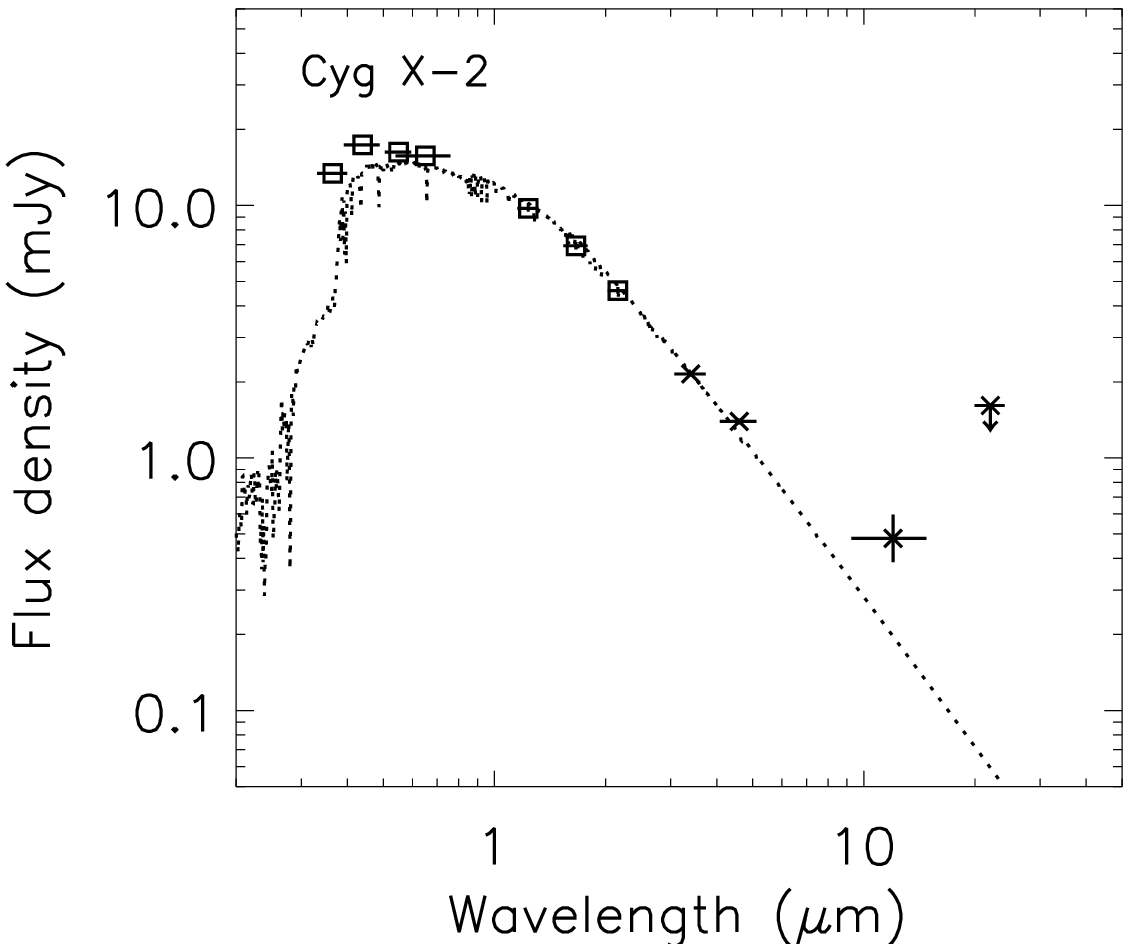}\\
\end{tabular}
\caption{Flux density spectra of Cen X-4, 4U~1700+24, 3A~1954+319, and
Cyg X-2. The squares are the optical/near-IR data points and crosses are
the WISE data points. The dotted curves represent emission from the companion
stars.
\label{fig:com}}
\end{center}
\end{figure*}

For A0620$-$00 and XTE J1118+480,
targeted \textit{Spitzer} imaging was carried out and
the detection and identification of circumbinary disks was reported by
\citet{mm06}. In Figure~\ref{fig:bd}, their dereddened broad-band optical
and IR flux density spectra are shown, 
and the \textit{Spitzer} flux measurements or upper limit are included
(diamonds in Figure~\ref{fig:bd}).  As can be seen, the WISE measurements
or upper limits are generally consistent with the \textit{Spitzer} data
(note that for faint, $\gtrsim 14$ mag sources, WISE measurements have
large uncertainties and
biases\footnote{see http://wise2.ipac.caltech.edu/docs/release/all sky/expsup/sec6\_3c.html},
which may be the cause for a brighter W2 flux detected from XTE J1118+480
than that at \textit{Spitzer} 4.5~$\mu$m band).

For A0620$-$00 and XTE J1118+480, we re-studied their debris disks with
updated information and optical and near-IR data. Using the spectral types,
estimated distances, and binary parameters published in literatures
(see Table~\ref{tab:prop}), a stellar spectrum plus that of a standard 
optically thick,
geometrically thin accretion disk \citep*{ss73,fkr02} were combined
to describe the observed optical and near-IR flux densities. For A0620$-$00,
the masses of the black hole primary and companion and thus the binary orbit's
inclination angle $i$ used were 11 and 0.7~$M_{\sun}$, 
and 41\arcdeg\ \citep{gho01},
respectively. For XTE J1118+480, they were 8.5 and 0.4 $M_{\sun}$, and
68\arcdeg\ \citep{gel+06}. For the black hole systems in quiescence, the
inner edge of a standard accretion disk is not certain (e.g., \citealt{yn14}),
but the uncertainty affects very little of optical fluxes considered here.
The outer edge of a disk was set at 90\% of the Roche lobe radius of
a primary, at which the disk should be tidally truncated \citep{fkr02}. The
only free parameter adjusted to describe the optical and near-IR data was
the mass accretion rate $\dot{M}$ in the disk. We found that
$\dot{M}\simeq 1.5\times 10^{14}$ g~s$^{-1}$ and 5.5$\times 10^{13}$ g~s$^{-1}$
for A0620$-$00 and XTE J1118+480, respectively, were required to produce
the model spectra displayed in Figure~\ref{fig:bd}. 
The values are approximately consistent with
those reported in \citet{mm06}. 

In order to fully describe the overall broad-band spectrum of either
A0620$-$00 or XTE J1118+480, a third component
is apparently needed for the excess IR emission seen in the both
systems. Arguments for supporting the existence of a circumbinary
disk were discussed by \citet{mm06}. Here following their work,
the same disk model, $T_{\rm disk}\sim T_{\ast}r^{-3/4}$ where $T_{\rm disk}$
is the disk temperature at radius $r$ and $T_{\ast}$ is the temperature of
a companion, was used to fit the excess emission. The inner disk radius
from the center of masses of a binary was set to be 1.7$a$ \citep*{tsd03,mm06},
where $a$ is the binary separation. To simplify the model, the inclination
angle of the disk was set to be aligned with that of the orbit, and
the only free parameter was the outer disk radius $r_{\rm o}$. We found that
$r_{\rm o}=2.3a$ and 4.5$a$ for A0620$-$00 and XTE J1118+480, respectively,
were required to fit the excess emission (see Figure~\ref{fig:bd}).

Having an orbital period of 1160.8 days, the identified M6III companion
in GX 1+4 can not fill its Roche lobe (\citealt{hin+06} and references therein).
This explains the lack of emission from an accretion disk in its broad-band
optical and near-IR spectrum. However the WISE fluxes clearly indicate
the excess emission from the source. Using the model above, in which
the neutron star primary and companion masses of 1.35 and 1.0 $M_{\sun}$,
respectively, and $i\simeq 70\arcdeg$ were
assumed \citep{hin+06}, a circumbinary disk with an outer radius of 2$a$
can provide an acceptable fit to the excess emission although the model fluxes
(blue dashed curve in Figure~\ref{fig:bd}) at W1 and W2 bands are slightly
lower than the observed. A smaller than 1.7$a$ inner disk radius is
required in order to eliminate the deviation. Since a mass-ejecting outflow
is expected from a giant star and optical spectroscopy has suggested
the presence of a $6\times 10^{13}$ cm radius gas envelope around
the binary \citep{dmb77}, we also tested to fit the excess emission with
a model of an optically thin dust shell (\citealt*{wcg88}; see also
\citealt{van+94}). We simplified the model by assuming the temperature
of 1200~K at the inner edge of the dust shell (the dust sublimation
temperature), the emission efficiency of dust grains $Q(\nu)\sim \nu^2$,
the number density distribution of the dust shell $n(r)\sim r^{-2}$, where
$r$ is the radius, and thus temperature function $T(r)\sim r^{-1/3}$.
The model emission from such a dust shell is displayed in Figure~\ref{fig:bd},
and it provides an acceptable description to the excess emission too and
is able to eliminate the deviation at W1 and W2 bands 
when the circumbinary disk model was considered.
The total number of dust grains in the shell is $\simeq$1.8$\times 10^{37}$, 
resulting in a mass of 10$^{-10}\ M_{\sun}$ when assuming 
3~g~cm$^{-3}$ for the density of grain material 
(grain radius of 0.1~$\mu$m was assumed). 
The mass value is a reasonable estimate for the dust around a giant star.

\subsection{Jet systems}

The neutron star LMXB 4U 0614+091 was detected by WISE at its first
three bands. This binary has been well studied from previous
targeted \textit{Spitzer} observations and its IR emission has been
identified to be from a jet
\citep{mig+06,mig+10}. The WISE fluxes at W1, W2, and W3 were
0.13$\pm$0.01, 0.15$\pm$0.02, and 0.61$\pm$0.14 mJy, respectively.
Given that the source was faint to WISE, we consider the WISE measurements
are consistent with that of \textit{Spitzer} (e.g., \citealt{mig+10}), although
flux variability is also expected for a jet. Since detailed, simultaneous
multi-band studies of the binary and its jet emission have been presented
by \citet{mig+10}, no further data analysis was conducted for the source.

The black hole binary GX 339$-$4 has long been known to have a jet
(e.g., \citealt{cf01} and references therein). Combining with 
quasi-simultaneous data of a large wavelength range from radio to X-ray, 
\citet{gan+11} have already analyzed the WISE data for the binary in detail 
and showed
that the WISE band emission was from a jet of the binary system. 

\subsection{Companion stars}

On the basis of their broad-band spectra, the companion stars in the LMXBs
Cen X-4, 4U 1700+24, 3A 1954+319, and Cyg X-2 were 
identified to have been detected by WISE, with no excess emission found at
the WISE bands. The broad-band spectra are shown in Figure~\ref{fig:com}. 
Among them, the W2 fluxes of 4U~1700+24 and 3A 1954+319 deviate
from the stellar model spectra significantly. However this is due to saturation
of the two sources in the WISE W1 and W2 images. They were very bright
($\sim 3$ mag; Table~\ref{tab:sum}), and although large flux uncertainties 
at the two bands (see Table~\ref{tab:sum}) reflects photometry of 
the saturated sources, 
the broad-band spectra indicate that their W2 fluxes were likely overestimated
by WISE photometry.

For 3A 1954+319, the companion star was classified as a M4-5III giant 
from optical spectroscopy by \citet{mas+06}. Using the $R$-band magnitude
given in the USNO-B1.0 catalogue \citep{usnob} and 
$V-R=1.58$ for an M4III star, \citet{mas+06} derived $V=10.7$ and thus estimated
a source distance of $\sim$1.7~kpc. We checked the USNO-B1.0 catalogue and
found that the only measurement for the source is $R=10.2$, which is 
not consistent with the value given in \citet{mas+06}. As can be seen in
Figure~\ref{fig:com}, the $R$-band flux is an order of magnitude lower than
the model spectrum, which may be due to large uncertainties in
photometric measurements in the USNO-B1.0 catalogue. The extinction
towards this high Galactic latitude ($Gb=64\arcdeg$) source is estimated
to be very low
($E(B-V)\simeq 0.03$; \citealt{sfd98}), hard to explain the discrepancy.
In any case, from matching an M4 giant model spectrum to the near-IR 2MASS and
WISE data, we found that the ratio of the stellar radius $R$ to distance $d$ 
is $R/d\simeq 2.7\times 10^{-9}$, and estimated $d\sim$0.67~kpc 
when $R=80\ R_{\sun}$.

For Cyg X-2, a spectral type of A9 for the companion was identified by
\citet{cck98}. We used a model spectrum of such a giant star to fit the
broad-band spectrum, and the model describes the observed IR data points well.
Since Cyg X-2 is a bright, persistent X-ray source, in which an
accretion disk exists, the optical fluxes contain emission
from the disk and thus are above the model spectrum. The W3 flux is
above the model spectrum, but we note that the measured magnitude, 11.95,
is close to the limiting magnitude of WISE at the W3 band \citep{wri+10} and
the deviation is likely due to the large uncertainties in the flux measurements
of very faint sources. The $R/d$ ratio we obtained was 2.5$\times 10^{-11}$.
Using the masses of 1.71 and 0.58 $M_{\sun}$ for the neutron star and 
companion \citep{cas+10}, respectively, the Roche lobe radius of 
the companion is
approximately 7.3~$R_{\sun}$, which implies a source distance of 6.7 kpc.
This distance value is consistent with 7.2$\pm$1.1~kpc, previously estimated 
by \citet{ok99}. 
\begin{figure}
\centering
\includegraphics[scale=0.58]{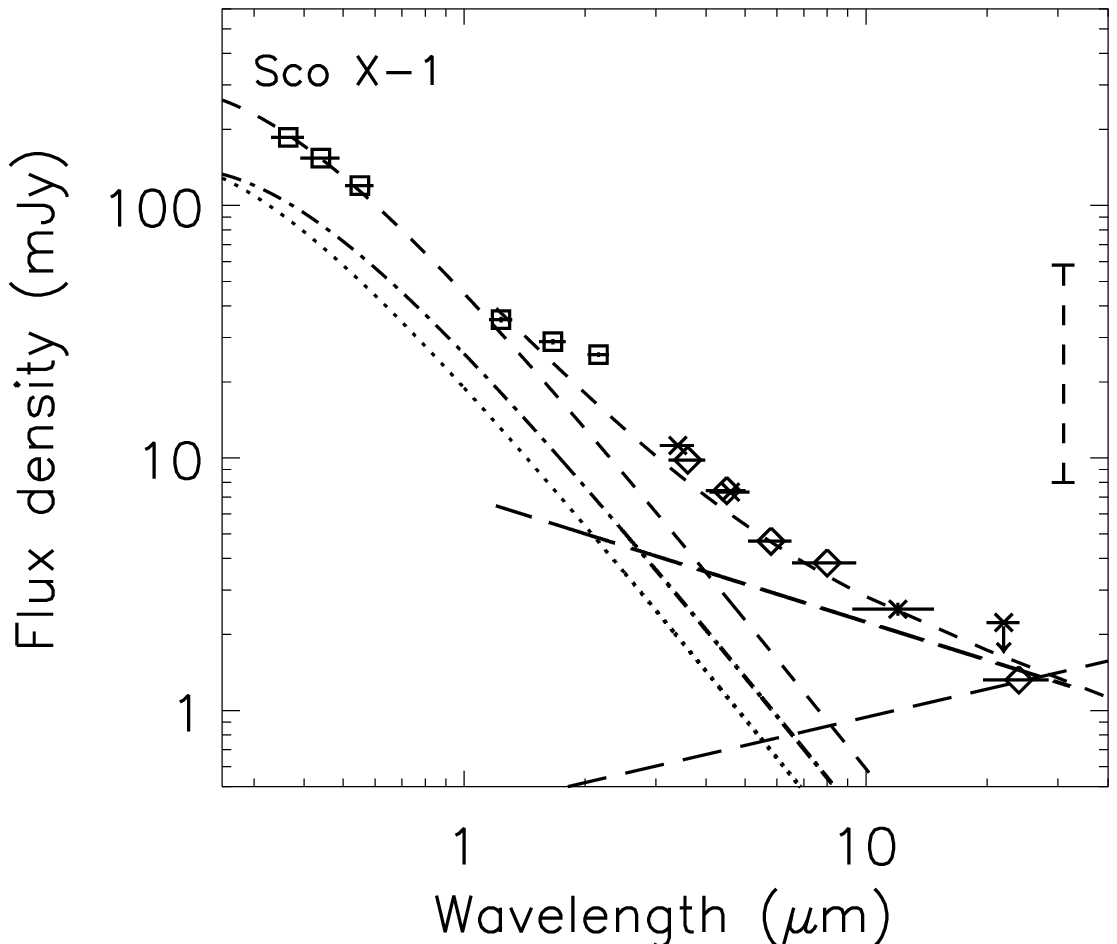}
\includegraphics[scale=0.58]{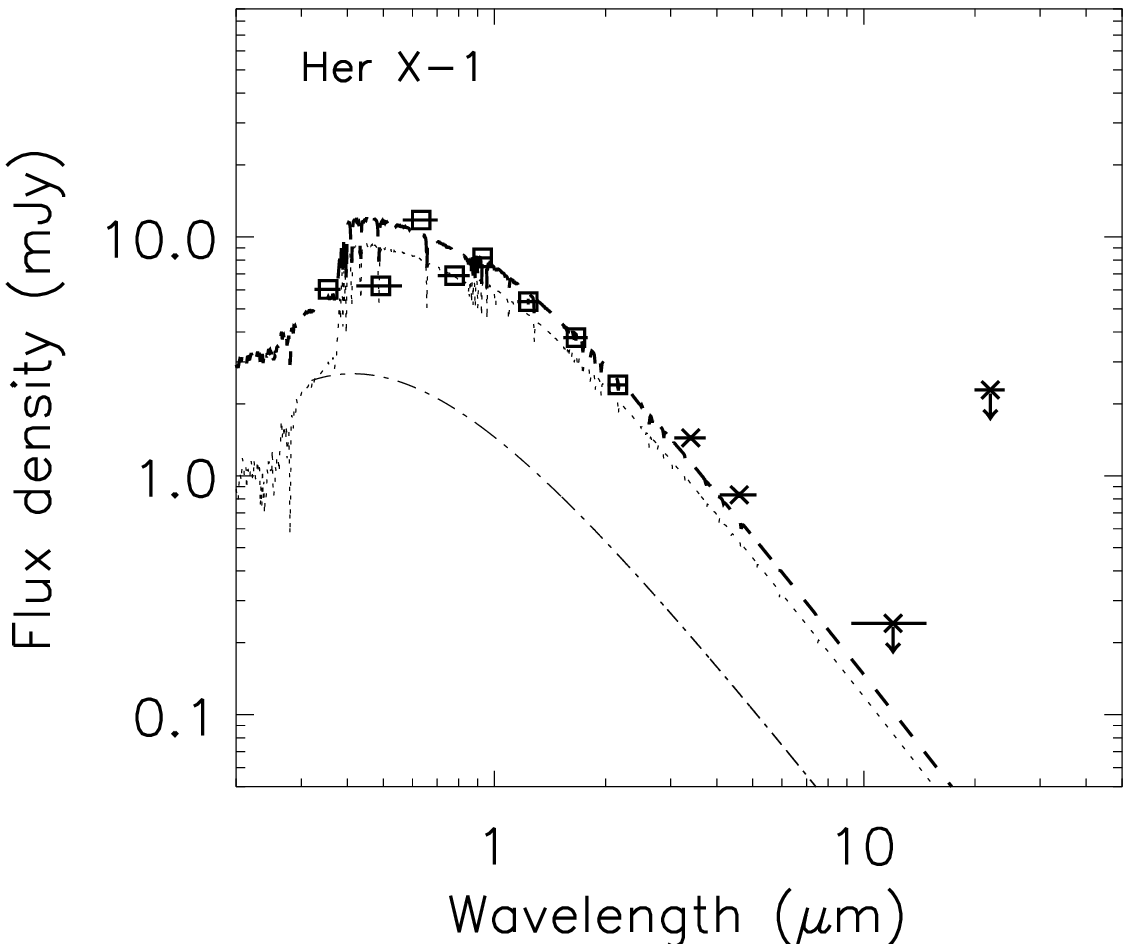}
\includegraphics[scale=0.58]{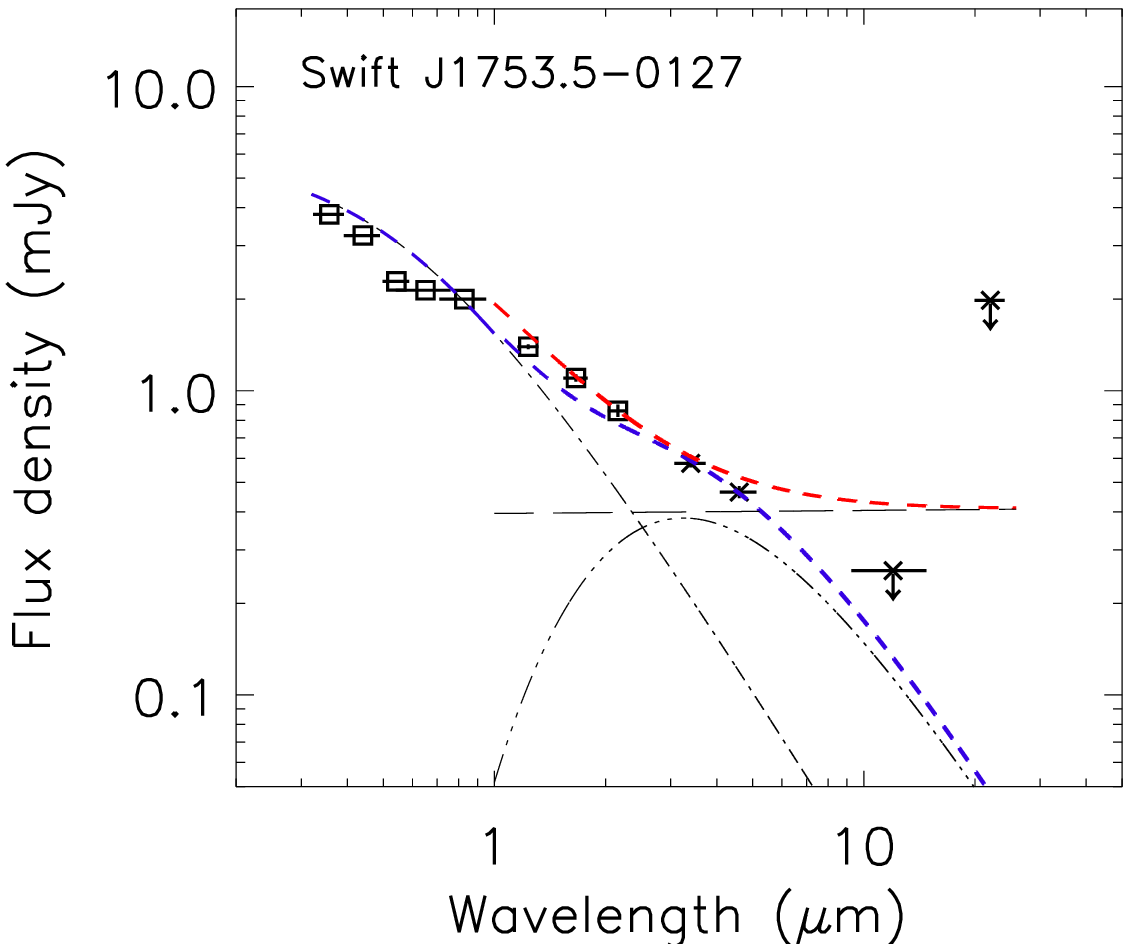}
\caption{Flux density spectra of Sco X-1, Her X-1, and Swift J1753.5$-$0127.
The squares and diamonds are optical/near-IR and \textit{Spitzer} data points,
respectively, and crosses are WISE data points. The dotted curves and
dash-dotted curves represent emission from the companions and accretion
disks, respectively, and the dashed curves represent the total emission
from all the components. For Sco X-1 and Swift J1753.5$-$0127, two
other components, the dash--triple-dotted and long dashed curves are 
discussed (see Section~\ref{subsec:pec}). The red and blue dashed curves
shown for the latter source are the total emission from a power-law and
a thermal component, respectively. The large error bar in the top 
(Sco X-1) panel indicates the range of the obtained radio fluxes 
in \citet{pan+07}.
\label{fig:pec}}
\end{figure}

\subsection{Peculiar systems}
\label{subsec:pec}

Two well known systems, Sco X-1 and Her X-1, were detected by the WISE
survey. These two neutron star LMXBs are very bright X-ray sources and
highly variable, due to accretion activity and orbital modulation.
Simultaneous, multiwavelength measurements are particularly required
for these two sources in order to conduct data analysis and to determine 
the presence of excess IR emission and the origin if the excesses are found. 
In Figure~\ref{fig:pec}, their broad-band spectra, which are not
simultaneous data, are shown. 

For Sco X-1, its companion star likely has a mass of 0.42 $M_{\sun}$ 
when a 1.4 $M_{\sun}$ neutron star is assumed 
($i\simeq 38\arcdeg$; \citealt{sc02}), 
and probably has a spectral type of earlier than G5 \citep{ban+99}. At
a distance of 2.8 kpc \citep{bfg99}, such a star has little contribution
to the optical fluxes of the source. However given the large X-ray luminosity
($L_{\rm X}\simeq 2\times 10^{38}$ erg s$^{-1}$; e.g., \citealt{bgf03}),  
we may estimate the irradiated temperature for the companion and it would
be of the order of 25,000--30,000 K, depending on the albedo of the companion
star. The model spectrum of this strongly irradiated star
and that of an irradiated accretion disk \citep{vrt+90} 
are shown in Figure~\ref{fig:pec}. The addition of the two spectra
is able to describe the optical and near-IR $J$-band data but 
significantly deviates from the WISE and \textit{Spitzer} data points
(the latter are from \citealt{rzm12}), suggesting the need
of an additional component. Given that Sco X-1 is known to have a radio jet
(e.g., \citealt{fgb01}),
we tested to add a power-law component to eliminate the deviation, and found
that $F_{\nu}\sim \nu^{0.5}$ could provide an acceptable fit. However,
the power-law component does not connect directly to the radio emission
from Sco X-1, which was found to be, for example, in a range 
of 8--59 mJy at 1.28 GHz \citep{pan+07}. A rising, $F_{\nu}\sim \nu^{-0.4}$ 
spectrum would be needed
in order to connect the \textit{Spitzer} 24~$\mu$m datum to the radio data.
We realized that since the data points shown in Figure~\ref{fig:pec} were
not simultaneously obtained, an alternative explanation for the discrepancy 
is that during either the WISE or \textit{Spitzer} observations, whose flux
measurements are consistent with each other, the radio emission would 
have been much faint if the IR emission had arisen from the jet.

For Her X-1, by studying its ultraviolet spectra, \citet{cvr95} found that
a model that contains
a companion of late A spectral type plus an X-ray irradiated disk could provide
a good fit. Here we used the binary parameters estimated by \citet{rey+97},
the masses of 1.5~$M_{\sun}$ and 2.3~$M_{\sun}$ for the neutron star and
companion, respectively, and found that the broad-band data basically could
be described by the model emission. Probably because our data were not 
simultaneous and the contribution from the companion was underestimated
(emission from a late type A star with a Roche-lobe radius of 
3.85 $R_{\sun}$ was used), the distance had to be 5.8 kpc, lower than
6.6$\pm$0.4~kpc estimated by \citet{rey+97}. In any case our analysis
shows that the IR data points are consistent with being
a Rayleigh-Jeans tail of thermal emission and no excess emission was found at 
the IR bands.
\begin{figure*}
\centering
\includegraphics[scale=0.58]{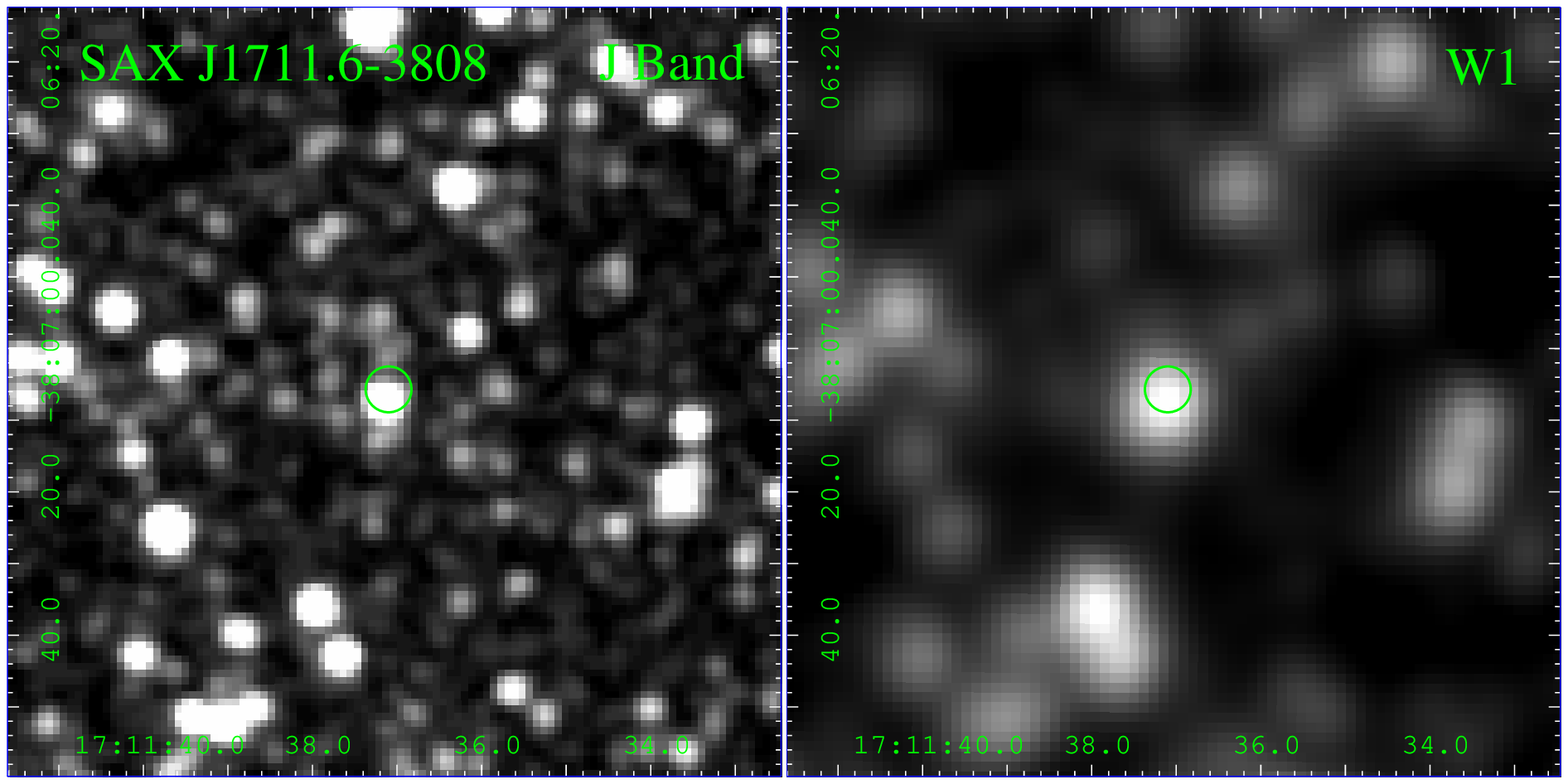}
\includegraphics[scale=0.58]{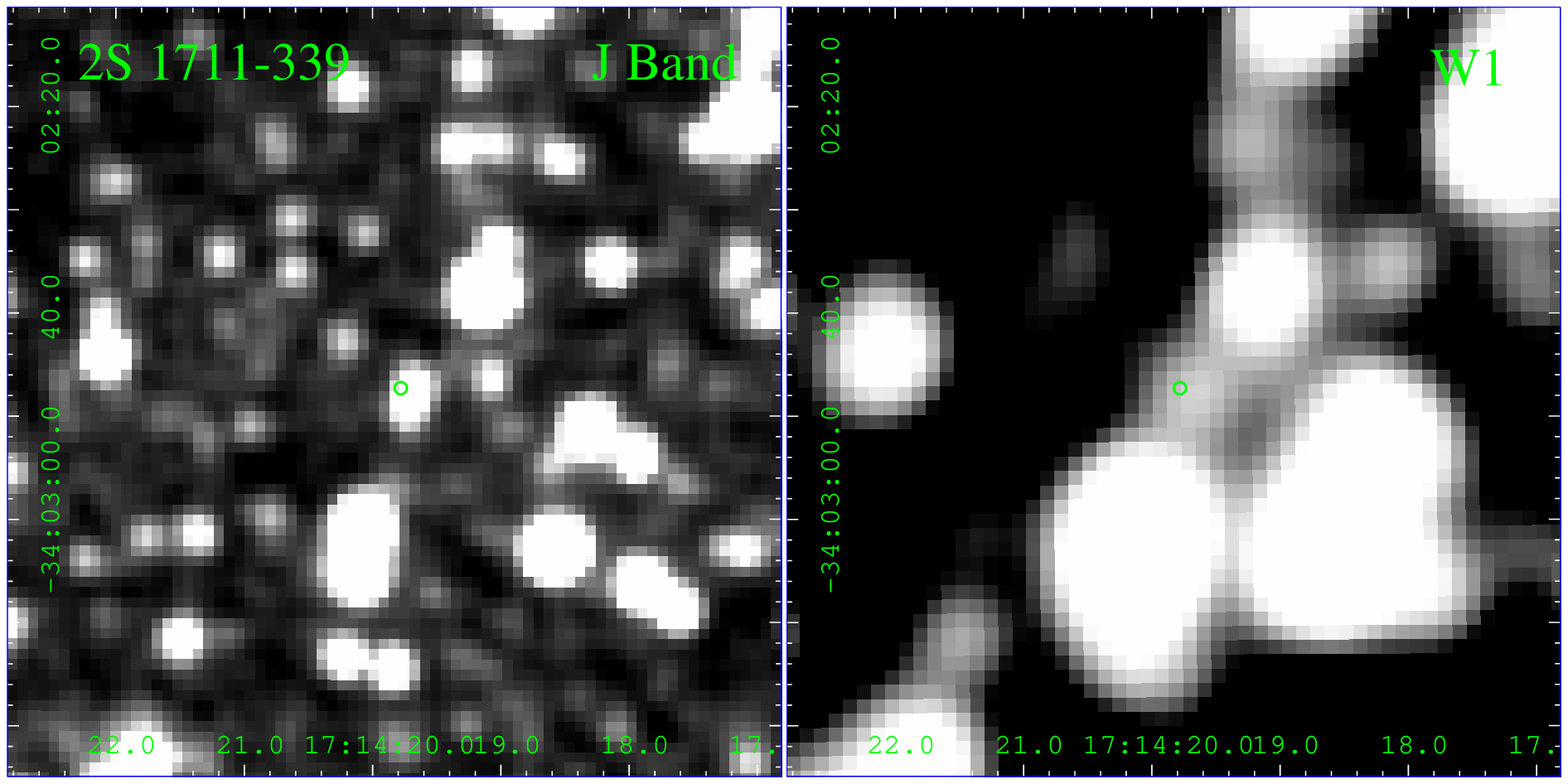}
\caption{Near-IR $J$ band (left panels) and WISE W1 band (right panels)
images of the fields of SAX J1711.6$-$3808 and 2S~1711$-$339. The X-ray 
positions are indicated by the circles in the centers of the images.
\label{fig:fv}}
\end{figure*}

The black hole candidate Swift J1753.5$-$0127 was discovered in 2005
\citep{pal+05}. Since then, many studies of the binary have been carried out
(e.g., \citealt{fro+14} and references therein). In Figure~\ref{fig:pec},
the broad-band optical and near-IR data from \citet{fro+14} are shown,
which were obtained in 2012 nearly simultaneously. The model fit, 
produced from a thin, steady-state accretion disk, is also from \citet{fro+14},
and can generally describe the source's emission in the range of
from ultraviolet to near-IR wavelengths. The deviation starts from near-IR
$H$ band, which was noted by \citet{fro+14}, and it is so significant at
the WISE bands that an additional component is clearly needed. The binary
is known to have a jet \citep{sol+10}. The radio observation that was
closest to the WISE observations was carried out in 2009 June, which found
a flux of 0.4 mJy at 8.4 GHz \citep{sol+10}. A power-law spectrum 
$F_{\nu}\sim \nu^{-0.01}$, presumably from the jet and connects to
the radio datum, can explain the IR excesses, but is not allowed by
the WISE W3 band flux upper limit. Again like the Sco X-1 case,
in order to connect to the radio data, a rising spectrum from mid-IR
(at the wavelength of $\sim$12 $\mu$m) is needed. Alternatively, a circumbinary
debris disk may also explain the IR excesses. However, since the companion
star was not seen in the emission from the binary, no constraints on
the disk temperature can be set. To produce the IR spectrum of the disk
shown in Figure~\ref{fig:pec}, the temperature at 1.7$a$ was 2500~K, too high
to be the inner edge of a dust disk. A slightly larger inner disk radius
is needed. In addition the outer radius was 
5.6$a$, larger than that in the A0620$-$00 and XTE J1118+480 cases.

\subsection{Candidate counterparts}

Little information is available for the two remaining sources:
SAX J1711.6$-$3808 and 2S 1711$-$339. The 2MASS $J$ band and WISE
W1 band images for each of them are shown in Figure~\ref{fig:fv}.
Given that the WISE candidate counterparts were detected by 2MASS, we further
compared their 2MASS positions with the reported X-ray positions, 
since the 2MASS
positions have a systematic uncertainty of only 0\farcs15, much more
accurate than that of the WISE. For SAX J1711.6$-$3808, the X-ray positional 
uncertainty is 3\farcs2, and from the 2MASS $J$-band image, it can be seen 
that it is located in a crowded field, indicating that the chance of 
coincidence is high. For 2S 1711$-$339, the \textit{Chandra} position with
an uncertainty of 0\farcs6 was reported \citep{wil+03,tor+04}.
The separation distance of the 2MASS source to the position is 1\farcs2,
approximately 2$\sigma$ away.
Given these, we only suggest the WISE sources as the candidate counterparts.

\section{Discussion}
\label{sec:dis}

Searching through WISE data, we have found 13 counterparts and 2 candidate 
counterparts among 187 LMXBs catalogued in \citet{lvv07}.
By collecting published results and/or analyzing the constructed broad-band 
spectra for the 13 counterparts, we have identified the origin of the
WISE-detected emission. Four of them probably have a candidate 
circumbinary disk, two harbor a jet, four had thermal emission from
their companion stars, and three are peculiar
systems with the origin of their IR emission more or less uncertain.
It should be noted that these LMXBs are highly variable sources, due
to accretion and/or related activities. 
For the two jet systems, simultaneous multiwavelength observations have 
been well carried out (e.g., \citealt{mig+10}), but for the other systems
the broad-band data used and analyzed in this work
are from different epochs, and therefore our studies of the
LMXBs are qualitative at most. In order to quantitatively constrain the 
properties of the candidate circumbinary disk systems or to identify 
the origin of the IR emission detected from the three peculiar systems, 
simultaneous observations at from X-ray, optical/IR, to radio frequencies
should be carried out.  
\begin{figure}
\begin{center}
\includegraphics[scale=0.58]{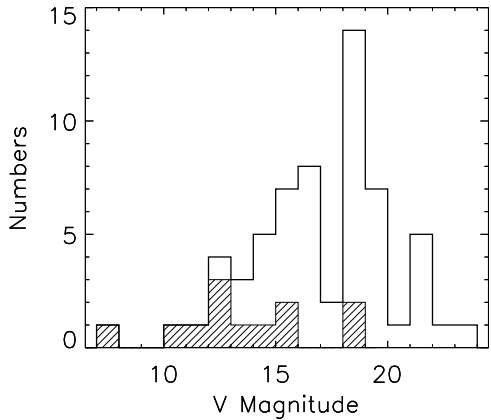}
\caption{$V$ magnitude distribution of 61 LMXBs that have reported 
$V$ measurements. The filled histogram marks that of the 12 LMXBs with the WISE 
counterparts. GRS 1915+105 is not included due to high extinction.
\label{fig:vdis}}
\end{center}
\end{figure}

For the candidate circumbinary disk systems, 
\citet{gal+07} have proposed the jet scenario to explain the excess IR emission
seen in A0620$-$00 and XTE J1118+480.
Given the ubiquity of the presence of jets in the LMXBs \citep{fen06}, 
the jet scenario is also plausible. Detection of dust signatures
such as the PAH emission features seen in GRS~1915+105 is needed in order
to verify the presence of a dust disk. On the other hand, the 
case of GRS~1915+105, a well-known microquasar, has shown that an LMXB can
harbor both a circumbinary disk and a jet. Thus the two binaries could also
have the both. For GX~1+4, although we can not determine if the excess emission
arises from a dust shell or a circumbinary disk, the WISE detection of 
the excesses has revealed another interesting aspect for this binary: 
the giant-star companion's significant mass loss is also observable at IR 
wavelengths.

Finally it is interesting to check the detectability of the LMXBs by 
an IR survey like WISE. In Figure~\ref{fig:vdis}, we show the numbers of 
the LMXBs as the function of their $V$ magnitude reported in \citet{lvv07}.
There are totally 61 LMXBs that have a reported $V$ magnitude value.
The distribution of the 12 LMXBs with the WISE counterparts, not including 
GRS 1915+105 due to extremely high extinction, is shown as the filled 
histogram.
It is not surprising that the WISE-detected sources are among the brightest.
The two exceptions are at 18--19 mag, which are 4U 0614+091 and GX 1+4,
the first due to the rising spectrum of its jet \citep{mig+06} 
and the latter due to high extinction (Table~\ref{tab:prop}). 
From the statistics based on this small sample, while the bias towards
the intrinsically bright X-ray sources (such as Sco X-1 and Her X-1) exists,
we may suspect that more than half of the LMXBs would have
non-stellar IR emission, due to the presence of a jet, a circumbinary
debris disk, or the both.

\acknowledgements

We thank Fuguo Xie for helpful discussion about jet systems
among the X-ray binaries.
The publication makes use of data products from the Wide-field
Infrared Survey Explorer, which is a joint project of the University
of California, Los Angeles, and the Jet Propulsion Laboratory/California
Institute of Technology, funded by NASA. This publication also makes use
of data products
from the Two Micron All Sky Survey, which is a joint project of
the University of Massachusetts and  the Infrared Processing and
Analysis Center/California Institute of Technology, funded by the National
Aeronautics and Space Administration and the National Science Foundation.

This research was supported by the National Natural Science Foundation of
China (11373055) and the Strategic Priority Research Program
``The Emergence of Cosmological Structures" of the Chinese Academy
of Sciences (Grant No. XDB09000000). Z.W. is a Research Fellow of the
One-Hundred-Talents project of Chinese Academy of Sciences.

\bibliographystyle{apj}

\begin{deluxetable}{l c c c c c c c c c}
\tabletypesize{\footnotesize}
\tablecaption{Summary of WISE measurements of the (candidate) counterparts of 
15 LMXBs\label{tab:sum}}
\tablewidth{0pt}
\tablehead{\colhead{Source} & \colhead{$\delta$}  &
\colhead{Obs. date}  & \colhead{W1}  & \colhead{W2}  & \colhead{W3\tablenotemark{a}}  &
\colhead{W4\tablenotemark{a}}  & \colhead{$J$}  & \colhead{$H$}  & \colhead{$K$}\\
\colhead{} & \colhead{(\arcsec)} & \colhead{} & \colhead{(mag)}
& \colhead{(mag)} & \colhead{(mag)} & \colhead{(mag)} & \colhead{(mag)} & \colhead{(mag)} & \colhead{(mag)}
}
\startdata
A0620$-$00 & 0.88 & 2010-03-19 & 14.17$\pm$0.03 & 13.84$\pm$0.04 & 12.71 & 9.11 & 15.49$\pm$0.05 & 14.74$\pm$0.06 & 14.38$\pm$0.07 \\
4U 0614+091 & 0.57 & 2010-03-18 & 16.02$\pm$0.09 & 15.19$\pm$0.14 & 11.75$\pm$0.25 & 8.78 &  &  & \\
XTE J1118+480  & 0.88 & 2010-05-10 & 16.40$\pm$0.08 & 15.85$\pm$0.17 & 12.92 & 9.24 & 12.76$\pm$0.02 & 12.44$\pm$0.03 & 12.08$\pm$0.02 \\
4U 1456$-$32 (Cen X-4) & 0.65 & 2010-02-05 & 14.65$\pm$0.04 & 14.66$\pm$0.08 & 12.62 & 9.04 & 15.60$\pm$0.06 & 15.05$\pm$0.07 & 14.66$\pm$0.08 \\
H 1617$-$155 (Sco X-1) & 0.25 & 2010-02-23 & 11.16$\pm$0.02 & 10.97$\pm$0.02 & 10.20$\pm$0.06 & 8.95 & 11.91$\pm$0.03 & 11.55$\pm$0.03 & 11.15$\pm$0.02 \\
2A 1655+353 (Her X-1)  & 0.19 & 2010-02-21 & 13.32$\pm$0.03 & 13.28$\pm$0.03 & 12.70 & 8.90 & 13.73$\pm$0.03 & 13.61$\pm$0.03 & 13.63$\pm$0.03 \\
4U 1659$-$487 (GX 339$-$4)  & 1.06 & 2010-03-10 & 9.62$\pm$0.03 & 8.77$\pm$0.02 & 6.82$\pm$0.02 & 5.17$\pm$0.04 & 15.91$\pm$0.14 & 15.40$\pm$0.15 & 14.97$\pm$0.14 \\
4U 1700+24 & 0.16 & 2010-02-27 & 3.04$\pm$0.22 & 2.43$\pm$0.11 & 3.01$\pm$0.03 & 2.85$\pm$0.03 & 4.17$\pm$0.20 & 3.32$\pm$0.19 & 2.99$\pm$0.23 \\
SAX J1711.6$-$3808 & 1.81 & 2010-03-10 & 8.74$\pm$0.02 & 8.66$\pm$0.02 & 8.35$\pm$0.05 & 7.60$\pm$0.18 & 12.46$\pm$0.03 & 10.43$\pm$0.03 & 9.43$\pm$0.03 \\
2S 1711$-$339 & 1.19 & 2010-03-11 & 12.37$\pm$0.04 & 12.54$\pm$0.06 & 9.13$\pm$0.08 & 8.15$\pm$0.32 & 14.34$\pm$0.04 & 13.19$\pm$0.03 & 12.86$\pm$0.04 \\
4U 1724$-$307 & 1.38 & 2010-03-13 & 7.05$\pm$0.02 & 6.99$\pm$0.02 & 7.46$\pm$0.04 & 6.38$\pm$0.09 & 10.33$\pm$0.10 & 9.09$\pm$0.09 & 8.56$\pm$0.07 \\
3A 1728$-$247 (GX 1+4) & 0.19 & 2010-03-13 & 7.48$\pm$0.03 & 7.24$\pm$0.02 & 6.77$\pm$0.02 & 6.33$\pm$0.06 & 10.10$\pm$0.02 & 8.70$\pm$0.05 & 7.98$\pm$0.02 \\
Swift J1753.5$-$0127 & 0.11 & 2010-03-17 & 14.31$\pm$0.03 & 13.91$\pm$0.05 & 12.64 & 9.05 &  &  &  \\
GRS 1915+105 & 0.11 & 2010-04-11 & 12.11$\pm$0.03 & 11.45$\pm$0.04 & 9.86$\pm$0.24 & 7.68$\pm$0.33 & 15.43 & 14.37$\pm$0.06 & 12.90$\pm$0.06 \\
3A 1954+319 & 0.98 & 2010-04-27 & 3.41$\pm$0.10 & 3.10$\pm$0.07 & 3.35$\pm$0.01 & 3.18$\pm$0.02 & 4.91$\pm$0.04 & 3.88$\pm$0.23 & 3.51$\pm$0.36 \\
4U 2142+38 (Cyg X-2) & 1.18 & 2010-06-03 & 12.89$\pm$0.02 & 12.72$\pm$0.03 & 11.95$\pm$0.24 & 9.28 & 13.40$\pm$0.03 & 13.16$\pm$0.04 & 13.05$\pm$0.03 \\
\enddata
\tablenotetext{a}{The magnitude values with no uncertainty given are upper limits.}
\end{deluxetable}

\begin{deluxetable}{l c c c c c c c c}
\tabletypesize{\footnotesize}
\tablecaption{Properties of the 15 LMXBs\label{tab:prop}}
\tablewidth{0pt}
\tablehead{\colhead{Name} & \colhead{Type} & \colhead{Optical} & 
\colhead{Spectral Type} & \colhead{$P_{\rm orb}$} & \colhead{Distance} &
\colhead{Optical data}  & \colhead{$E(B-V)$} & \colhead{References} \\
\colhead{} & \colhead{} & \colhead{} & \colhead{} & \colhead{(hr)} & 
\colhead{(kpc)} & \colhead{}  & \colhead{(mag)} & \colhead{}
}
\startdata
A0620$-$00 & BH & V616 Mon & K4V & 7.75 & 1.16$\pm$0.11 & $BVRI$ & 0.39 & 1,2,3 \\
4U 0614+091 & NS & V1055 Ori & & & 1.5$\sim$3 & $UBV$ & 0.3 & 4,5,6\\
XTE J1118+480 & BH & KV UMa & K5V$\sim$K7V & 4.08 & 1.8$\pm$0.60 & $R$ & 0.013 & 7,8,9,10,11 \\
4U 1456$-$32 (Cen X-4) & NS & V822 Cen & K5V & 15.1 & 1.2$\pm$0.3 & $BVRI$ & 0.1 & 12,13,14 \\
H 1617$-$155 (Sco X-1) & NS & V818 Sco & & 18.9 & 2.8$\pm$0.3 & $UBV$ & 0.3 & 15,16,17 \\
2A 1655+353 (Her X-1) & NS & HZ Her & & 40.8 & 6.6$\pm$0.4 & $u'g'r'i'z'$ & 0.05 & 18,19,20,21 \\
4U 1659$-$487 (GX 339$-$4) & BH & V821 Ara & & 42.14 & 8$\sim$2 & $BV$ & 1.1 & 22,23,24,25 \\
4U 1700+24 & & HD 154791 & M2III & 404d & 0.42$\sim$0.04 & $UBV$ & 0.044 & 26,27,28 \\
SAX J1711.6$-$3808 & BH & & & & & & 5 & 29 \\
2S 1711$-$339 & NS? & & & & & $R$ & & 30 \\
3A 1728$-$247 (GX 1+4) & NS & V2116 Oph & M5III & 1160.8d & 4.3 & $BVR$ & 1.63 & 31,32 \\
Swift J1753.5$-$0127 & BH & & & & 6 & $BVRI$ & 0.36 & 33,34 \\
GRS 1915+105 & BH & V1487 Aql & & 804 & 9.0$\pm$3.0 &  & 8.6 & 35,36,37 \\
3A 1954+319 & NS? & & M4III$\sim$M5III & & 1.7 & $V$ &  & 38 \\
4U 2142+38 (Cyg X-2) & NS & V1341 Cyg & A9III & 236.2 & 7.2$\pm$1.1 & $UBVR$ & 0.40 & 39,40,41,42 \\
\enddata
\tablerefs{
(1) \citet{mr86};
(2) \citet{gho01};
(3) \citet{wu+76};
(4) \citet{pae+01};
(5) \citet{dav+74};
(6) \citet{rus+07};
(7) \citet{uem+00};
(8) \citet{mcc+01a};
(9) \citet{zur+02};
(10) \citet{hyn+00};
(11) \citet{mcc+03};
(12) \citet{che+89};
(13) \citet{snc93};
(14) \citet{bla+84};
(15) \citet{gwl75};
(16) \citet{bfg99};
(17) \citet{vrt+91};
(18) \citet{tan+72};
(19) \citet{rey+97};
(20) \citet{aba+09};
(21) \citet{bor+97};
(22) \citet{hyn+03};
(23) \citet{zdz+04};
(24) \citet{kon+00};
(25) \citet{bv03};
(26) \citet{gsk02};
(27) \citet{mas+02};
(28) \citet{gar+83};
(29) \citet{int+02};
(30) \citet{tor+04};
(31) \citet{hin+06};
(32) \citet{cr97};
(33) \citet{cad+07};
(34) \citet{sti+05};
(35) \citet{gcm01};
(36) \citet{bgm96};
(37) \citet{cc04};
(38) \citet{mas+06};
(39) \citet{cch79};
(40) \citet{cck98};
(41) \citet{ok99};
(42) \citet{mcc+84};
}
\end{deluxetable}

\end{document}